# UK White Paper on Space-based total solar eclipse observations: structure and dynamics of the solar atmosphere

Submitted to UK Space Frontiers 2035

Thematic area: Sun and heliosphere

Lucie Green (lucie.green@ucl.ac.uk), on behalf of co-authors and signatories.

## Executive summary

Our Sun is uniquely placed to enable a detailed study of astrophysical plasmas and how they are governed by the magnetic fields that thread through them. On the one hand, magnetic fields confine plasma and determine plasma heating, flows, and energisation. On the other hand, magnetic fields and their evolution give rise to the most violent eruptions in the Solar System. Understanding the details of how energy is built up and released, and the impact of these physical processes on the plasma, remain key open questions that directly map to UKRI's science strategy through the STFC Solar System Advisory Panel's roadmap for Solar System research goals: What are the causes, consequences & predictability of solar magnetic variability & the solar cycle? What are the structures, dynamics & energetics of the Sun? What are the underlying processes that drive Sun-planet connections? And what are the fundamental processes at work in the Solar System?

As laid out in this White Paper, the Moon-Enabled Sun Occultation Mission (MESOM) directly addresses these questions and in doing so delivers several "Pillars" of the National Space Strategy. 1: Unlocking growth in the UK space sector by supporting a whole-UK space ecosystem and developing nation-wide skills in building, operating, and navigating spacecraft beyond low-Earth orbit. 2: Collaborating internationally through our UK-Europe-US-Australia consortium. 3: Growing the UK as a science and technology superpower. The data provided by MESOM will also address many of the science questions detailed in the following UK Space Frontiers White Papers: Magnetic reconnection; Understanding solar flares and energetic events; Understanding physical processes in the middle corona; Measuring the coronal magnetic field with NIR spectropolarimetry. The work proposed in the Numerical modelling for solar and space applications White Paper will underpin, and help ensure high scientific return, for MESOM.

## 1. Scientific questions and general observation strategy

MESOM aims to address several high-priority science questions related to the solar atmosphere and its links to the solar wind. To achieve this, it is necessary to make quantitative determinations of the properties of both the plasma and magnetic field spanning the chromosphere, to the corona and into the region at the base of the solar wind. By flying a spacecraft near the apex of the Moon's umbral region and therefore creating total solar eclipses in space once every synodic month (~29.6 days), MESOM will provide observations of unprecedented resolution and quality to answer our key science questions.

The UK is at the forefront of eclipse experimentation and scientific studies of the solar atmosphere across several key areas. Dedicated ground-based UK eclipse experiments, providing spectroscopic and polarisation brightness data, reveal physical properties of the magnetised corona (e.g. Muro et al., 2023; Edwards et al., 2023). More broadly, use of both ground-based and space-based data from a wide range of instruments has led to world-leading expertise in the UK in mapping the plasma structure, composition, temperature and density diagnostics to test coronal heating theories, how energy and momentum is transported through the corona and to reveal source regions of the solar wind (e.g. Morton et al. 2015; Mihailescu et al., 2022; Baker et al., 2023). Within the coronal magnetic field, magnetic waves transport and deposit energy. Research in the UK has recently provided the first direct evidence of small-scale torsional Alfvén waves in the Sun's corona providing a means to validate the range of theoretical models that describe how Alfvén wave turbulence powers the solar atmosphere (Morton et al., 2025). The UK is also at the forefront of modelling the Sun's large-scale magnetic field, its magnetic topology and reconfiguration as a result of magnetic reconnection (e.g. Yeates et al., 2018). Magnetic reconnection has also been shown to play a key role in building structures that later erupt in events known as coronal mass ejections (e.g. James et al., 2020, for a review see Green et al., 2018), and interchange reconnection and dynamic activity that provide the necessary conditions for the escape of plasma into the solar wind (Owens et al., 2020; Ngampoopun et al., 2025). We describe here the UK priorities in these science areas, how this builds on our world-leading expertise, and outline observational requirements.

**What is the strength and structure of the global magnetic field of the Sun?** Knowledge of the global solar magnetic field is currently obtained from cutting edge modelling approaches. However, these models suffer from limitations in their underlying assumptions and require significantly improved observational validation to provide additional constraints. Such validation paves the way for future accurate time-dependent coronal magnetic field predictions, which then enable much-needed science and space-weather forecasting. For example, determining the large-scale organisation of the magnetic field and its hemispheric asymmetries provide constraints on how the dynamo processes generate and evolve field components, and global magnetic field models are a required input for solar wind and CME propagation models that are used for space weather forecasting purposes.

*Spectropolarimetric observations of the near-IR Fe XIII 1074 nm emission line from the upper chromosphere out to ~2* $R_s$ *will provide the coronal magnetic field vector required to determine the magnitude and global structure of the coronal magnetic field. High-resolution visible light broadband images covering the same region and with a spatial resolution on the order of 2", including polarised brightness measurements, will provide further context and constraints since the structure of the separatrix web is closely related to coronal plasma density. Combined, these spectropolarimetric and broadband observations will enable the validation of coronal models.*

**What is the role of magnetic waves in heating and accelerating plasma?** Alfvénic waves are a fundamental mechanism by which energy and momentum is transported through the solar atmosphere and into the solar wind. Although ubiquitous in the Sun's atmosphere, key questions remain. Notably, it is currently challenging to estimate the total wave energy flux, which is a vital constraint for the state-of-the-art Alfvénic wave turbulence models that can forecast basal estimates for coronal radiative output and solar wind mass flows. We also have limited knowledge on the partition of the wave energy flux resulting from different wave drivers, whether regional variations of the wave flux are present, and the associated wave energy deposition mechanisms (e.g., turbulence, instabilities, mode coupling). Hence, joint observations of time-series of the Alfvénic waves (imaging and Doppler) with simultaneous estimates for the local mean magnetic field and plasma density are required.

*Broadband visible light images of the solar atmosphere with very high spatial resolution (0.24", or 170 km at the Sun) and high cadence (4 seconds) over a small field of view (e.g. 0.5 x 0.5 $R_s$) will provide quantitative data on the plane-of-sky amplitude, periodicities, and propagation of Alfvénic fluctuations. Simultaneous narrowband spectral imaging of the Fe XI 637nm emission line at high temporal cadence will provide line-of-sight Doppler and linewidth information as well as non-thermal linewidths. Combined, these datasets enable detailed studies of waves, and their origin from upper chromosphere, transition region, and to the corona. This can be connected to the general plasma properties of the same regions through density estimates (via polarized brightness), and temperature/composition diagnostics via spectral imaging (next paragraph).*

**How and why does the plasma composition, density and temperature vary across coronal structures?** Temporal and spatial variations have been observed in the properties of the coronal plasma, which can be driven by various processes. One key varying property is composition: the action of the ponderomotive force that arises from the reflection or refraction of Alfvén waves in the chromosphere acts to transport elements with a low first ionisation potential (FIP) into the atmosphere above. Gravitational settling (or elemental settling) takes place higher in the corona and separates elements based on their mass, with higher mass elements settling towards the surface. In addition, magnetic reconnection between structures will lead to mixing of plasma. Measuring and tracking composition variation reveals a fundamental aspect of the solar plasma. Estimates of plasma density and temperature provide further information that, combined with composition, provide a way to probe energy and mass transport in the corona and beyond into the solar wind.

*Spectral imaging of multiple emission lines in the visible and NIR, covering elements with low and high FIP, with high signal to noise across the upper chromosphere to the corona is required to map the spatial distribution of elemental abundances, plasma density, temperature, and non-thermal broadening. This is combined with contextual information from a broadband visible imager, as well as an independent measure of plasma density through visible light polarized brightness.*

**What is the role of magnetic reconnection in restructuring the global corona from small to large scales, and how does it drive changes in the properties of the escaping plasma?** The corona is host to a wide variety of propagating small-scale events (e.g., Sheeley blobs, inflows, periodic density structures), all of which are believed to be manifestations of magnetic reconnection. Observations of their formation and early evolution

are rare and restricted to larger/brighter events, and their correspondence to the properties of the escaping plasma are indirect. At larger scales, the corona is restructured by coronal mass ejections (CMEs) that can disrupt streamers and produce extended current sheets, turbulence, flows and waves in their wake. The post-CME current sheet can last for days, and the expansion of the CME itself drives reconnection with surrounding fields (e.g., Bemporad et al., 2008). The redistribution of open flux caused by CMEs will have a corresponding impact on solar wind outflows, which is not yet well understood. Whether CMEs play a significant role in opening magnetic flux to the heliosphere is will be probed using observations that map the magnetic field configuration out to several solar radii combined with modelling of the large-scale magnetic field.

*High-resolution and high-cadence visible light broadband images of the middle corona and solar wind acceleration region will provide the ability to relate small-scale structures and dynamics to the global appearance of the corona and its connection to the heliosphere via the nascent solar wind, and will provide the data needed to understand how the inner heliosphere evolves due to open field from coronal holes, closed field of active regions, and dynamic events such as coronal mass ejections. The properties of the escaping plasma will be probed via simultaneous high quality spectral imaging in several emission lines.*

**What is the fine-scale structure of the extended solar wind, and how is this linked to small-scale events near the Sun?** In situ observations from Solar Orbiter and Parker Solar Probe have revealed the complexity and small-scale structure of the solar wind. There are open questions as to whether such small-scale structures, such as blobs and switchbacks, have a solar or a heliospheric origin, and the physical processes by which these structures form. Current gaps in our understanding of small-scale coronal and heliospheric processes hinder the development of more accurate background solar-wind models that are required for our scientific understanding as well as space weather forecasting and require high-resolution images of the density features in the corona to be mapped.

*High-resolution (2.2”) and high-cadence (15s) broadband heliospheric imager-mode data of the extended corona (above an altitude of $1R_s$ and into the solar wind region), with determination of plasma density through visible light polarized brightness.*

**Serendipity science: What are the processes leading to the destruction of Kreutz-group comets near the Sun, and what are the non-gravitational forces operating on comets?** Images of the occulted Sun and observations from heliospheric imagers remain the primary source of information on comets. The overwhelming majority of comets observed by coronagraphs are in the process of being actively destroyed, but the specific mechanisms and processes culminating in their destruction remain elusive. Observations of the fragmentation and deviation from gravitational orbits would enable break-up processes and sublimation-driven forces to be understood. Incoming sun-grazing comets are observed at a rate of approximately one comet every two to three days.

*High-resolution visible light broadband images centered near the sodium D emission lines and within 7 $R_s$ will enable the fragmentation of Kreutz-group comets and sungrazers to be determined and monitored. Additional spectroscopic observations will provide novel insight into the cometary composition and outgassing.*

## 3. Enabling mission: MESOM

MESOM (Moon Enabled Sun Occultation Mission) will view the atmosphere of the Sun from the upper chromosphere to the corona with total solar eclipse conditions, and higher altitudes of the solar atmosphere/solar wind acceleration region outside times of totality. MESOM will recreate total solar eclipse conditions in space by using the Moon as a natural occulter once every synodic month (~29.6). The concept, pioneered in the UK by Eckersley and Kemble (2017), greatly simplifies observations of the occulted Sun as no internal/external occulters are required, as is the case with the recently launched Proba-3 formation flying mission (Zhukov et al. 2025). MESOM capitalises on the mutual gravitational attractions of the Sun, Earth, and Moon to fly a 2:1 synodic resonant orbit that completes two revolutions about the Earth-Moon barycenter in one synodic month. The orbit yields more than 400 minutes of totality over two years of nominal science operations in which measurements can be taken free from the artifacts caused by occulting components, and the degradations caused by the Earth's atmosphere. Outside times of totality, MESOM observes the extended corona via instrumentation in 'heliospheric imager' mode. The orbit was identified, and the mission concept studied, in work funded by the UK Space Agency under the National Science Technology and Science and Exploration Bilateral Programmes (Baresi et al., 2024).

MESOM's suite of instruments exploits these optimal viewing conditions to enable quantitative measurements of coronal magnetic fields and plasma properties down to small spatial scales at high cadence. The payload is designed to provide co-ordinated observations that address the science questions discussed in Section 1. The large spacecraft-Moon distance avoids the diffraction/stray light issues of current space-based coronagraphs, hugely simplifying the design of MESOM's four complimentary instruments, which combined have a payload mass of less than 40 kg (MagCHILS 10 kg, HiBri 10 kg, LoBri 10 kg, CHILS 6 kg), and view the corona down to 1.02 $R_s$ or lower.

**1) MagCHILS** (Magnetic-Coronal High-Resolution Imaging Line Spectrometer) is an imaging spectropolarimeter of the Fe XIII 1074 nm emission line with a field of view up to 6 x 6 $R_s$ and 10" resolution. This crucial diagnostic is an entirely novel measurement from space which avoids the inherent difficulties of ground-based measurements of polarization. MagCHILS has heritage from Solar Orbiter/PHI, Sunrise and Vigil/PMI.

**2) HiBri** (High-resolution Broadband Imager) will observe visible light through a broadband filter of FWHM 100nm centered at 580nm across a field of view of 0.5 x 0.5Rs, pointing at/off the solar limb. A three-mirror anastigma telescope with 0.6m aperture allows unprecedented high spatiotemporal resolution of 0.24″ (170 km at the Sun) and 4s cadence. HiBri's spatial resolution surpasses any other instrument (visible light or otherwise) to observe the extended solar atmosphere, and builds on a heritage of total eclipse ground-based imagers.

**3) LoBri** (Low-resolution Broadband Imager) is a larger field-of-view (4.2 x 4.2 $R_s$) telescope observing broadband visible light to determine the magnetic connectivity from the chromosphere to the middle corona and context to other instruments. LoBri is labelled 'Low-resolution' to distinguish from HiBri, yet it has a spatiotemporal resolution of 2.2″ (1400 km at the Sun) and 15s cadence (total brightness) that surpasses all current space-based coronagraphs . A filter wheel will enable total brightness measurements, and polarizers at -60, 0, +60 degree for calculating polarized brightness for density diagnostics. Outside of eclipse totality, LoBri will use a baffle system and off pointing to provide high-resolution observations of regions of the extended corona and solar wind, essentially operating as a novel heliospheric imager. The baffles build on a heritage of heliospheric imagers including Parker Solar Probe WISPR and Solar Orbiter SoloHI.

**4) CHILS** (Coronal High-resolution Imaging Line Spectrometer) is a spectral imager that will map key plasma diagnostics across a 3.2 x 3.2 $R_s$ field of view using four emission lines (Fe X 789 nm, Fe XI 637 nm, Fe XIV 530 nm, Ar X 554 nm) once per eclipse period. Operating in a higher temporal-resolution mode of 12s, CHILS will also provide a time series of spectral information in a single emission line, which is necessary to capture small-scale motions and waves. The CHILS instrument design has been verified by industrial experts Qioptiq in a recent UK Space Agency project and is based on an STFC-funded spectrometer designed and built at Aberystwyth University (2016-19; ST/N002962/1), which has been deployed successfully at 3 total solar eclipses.

# 4. Global and strategic context

MESOM builds on a decades-long development of UK ground-based eclipse instrumentation and expertise, and decades of experience of space instrumentation, orbit design and small satellite development. MESOM's ability to observe the full height range in the solar atmosphere will enable energy flow/transport throughout the atmosphere and the creation of eruptive structures to be studied in a connected way. MESOM data will enable the UK to continue their world-leading work that probes the origins of the solar wind, a research area that has accelerated in the Solar Orbiter era. The improvement in global modelling enabled by MESOM will have application across UK solar physics research groups. The provision of novel datasets, such as studies of comets (e.g., composition), links to UK-led mission concepts such as Comet Interceptor and will be important for UK plasma and planetary scientists working in this area.The UK-led MESOM concept study was carried out by an international consortium between the UK and the USA, in partnership with scientists and engineers from Australia. Further development involved a consortium covering the UK, Europe, USA and Australia and led to a submission to ESA in response to their call for F-class missions. Future development of the MESOM concept will include low-thrust orbit design and transfer scenarios, and the development of the UK-led instrument CHILS. Through the existing international team it is anticipated that a Spanish consortium would lead MagCHILS (given their expertise from Solar Orbiter/PHI, Sunrise and Vigil/PMI) and the US would lead LoBri/Hri (given their decades long heritage in the provision of coronagraphs and heliospheric imagers). MESOM instrumentation will also be applicable to other mission opportunities that use lunar or artificial occultation of the Sun.

MESOM contributions to space weather forecasting will improve the accuracy and reliability of operational forecasts, which require accurate models of the solar coronal magnetic field. Knowledge exchange and model validation related to space weather forecasting will be enabled across three continents through our Europe-US-Australian international collaboration. We also plan to utilise the long-demonstrated inspiration of total solar eclipse to engage diverse communities. Our consortium has significant expertise in solar eclipse and comet (seen when the Sun is occulted) citizen science projects (e.g., the recent Citizen CATE 2024, and Sungrazer from 2003 to present), school engagement activities, artist collaborations, engagement with local astronomical societies and amateur radio societies, and various forms of social and traditional media.

## Authors and signatories

Lucie Green, UCL Mullard Space Science Laboratory, UK

Nicola Baresi, Surrey Space Centre, University of Surrey, UK

Huw Morgan, Aberystwyth University, UK

Matt Gunn, Aberystwyth University, UK

Amir Caspi, Southwest Research Institute, USA

Daniel B. Seaton, Southwest Research Institute, USA

Giulio Del Zanna, University of Cambridge and University of Leicester, UK

Yeimy J. Rivera, Center for Astrophysics | Harvard & Smithsonian, USA

Francisco Javier Bailén, Instituto de Astrofísica de Andalucía | CSIC, Spain

David Orozco Suárez, Instituto de Astrofísica de Andalucía | CSIC, Spain

David M. Long, Dublin City University, Ireland

Anthony R. Yeates, Durham University, UK

Peter Wyper, Durham University, UK

Nawin Ngampoopun, Max Planck Institute for Solar System Research, Germany

Nathalia Alzate, ADNET Systems, Inc., NASA GSFC, USA

Simone Di Matteo, The Catholic University of America, NASA GSFC, USA
Malcolm Druett, University of Sheffield, UK
Paolo Romano, INAF - Catania Astrophysical Observatory, Italy
Erika Palmerio, Predictive Science Inc., USA
Steve Eckersley, SSTL, UK
Stephanie Yardley, Northumbria University, UK (MSSL, DIPC)
Richard J. Morton, Northumbria University, UK
Eleanna Asvestari, University of Helsinki, Finland
Alexander James, UCL Mullard Space Science Laboratory, UK
Sarah Matthews, UCL Mullard Space Science Laboratory, UK
Manuela Temmer, Institute of Physics, University of Graz, Austria
Greta Cappello, Institute of Physics, University of Graz, Austria
Teodora Mihailescu, INAF Osservatorio Astronomico di Roma, Italy
Deborah Baker, UCL Mullard Space Science Laboratory, UK
Natalia Zambrana Prado, UCL Mullard Space Science Laboratory, UK
Hamish Reid, UCL Mullard Space Science Laboratory, UK
Alexander James, UCL Mullard Space Science Laboratory, UK
Dan Ryan, UCL Mullard Space Science Laboratory, UK
James McKevitt, UCL Mullard Space Science Laboratory, UK
Samuel Skirvin, Northumbria University, UK
Sargam Mulay, University of Glasgow, UK
Viktor Fedun, Sheffield University
Jinge Zhang, Paris Observatory - PSL
David H. Brooks, Computational Physics, Inc., USA
Hidetaka Kuniyoshi, Northumbria University, UK
Marianna Korsos, University of Sheffield
Robertus Erdelyi, University of Sheffield
Jackie Davies, UKRI STFC RAL Space
Mario Bisi, UKRI STFC RAL Space
Anshu Kumari, Physical Research Laboratory, India
Divya Paliwal, Physical Research Laboratory, India
Eamon Scullion, Northumbria University, UK
Alexander Russell, University of St. Andrews, UK
Gert Botha Northumbria University, UK
Abhirup Datta, Indian Institute of Technology Indore
Karl Battams, NRL, USA